\documentclass[10pt,letterpaper]{article}
\usepackage{opex3}
\usepackage{graphicx}
\usepackage{amssymb}

\newcommand\vect[1]{\mathbf{#1}}
\begin{document}

\title{Single-particle motional oscillator powered by laser}

\author{A. E. Kaplan}
\address{Dept. Electr. and Comp. Engineering,
Johns Hopkins University, Baltimore, MD 21218}
\email{alexander.kaplan@jhu.edu}

\begin{abstract}
An ion, atom, molecule or macro-particle in a trap
can exhibit large motional oscillations
due to the Doppler-affected radiation pressure by
a laser, blue-detuned from an absorption line of a particle.
This oscillator can be nearly thresholdless, but
under certain conditions it may exhibit huge hysteretic excitation.
Feasible applications include a "Foucault pendulum" in a trap,
a rotation sensor, single atom spectroscopy, isotope separation, etc.
\end{abstract}

\ocis{(140.7010) Laser trapping; (350.4855) Optical tweezers or optical manipulation}

\section{Introduction}
Self-sustained oscillators (SSO) [1]
are abundant in the world:
organ and violin, sea waves,
many biological processes, car engines,
clocks and watches, van-der-Pol oscillator, maser and laser,
El Ni$\tilde{n}$o and La Ni$\tilde{n}$a cycles,
the luminosity oscillations of many stars,
and perhaps even the universe itself, to name just a few.
While the energy supply (pumping)
can be provided by all kinds of sources,
the common facilitator of all SSO
is a so called positive feedback, which
overcomes damping by properly controlling
the system during an oscillation cycle.
A zero steady-state then becomes
unstable, and the oscillations grow until
they reach a stable limit cycle with a
well defined amplitude and frequency of coherent oscillation.
Many SSO also exhibit hysteretic excitation
(bistability) and related multi-limit cycles
due to nonlinearity.

It is of great interest to explore SSO in 
its most fundamental setting by using single particles.
Amazingly, the effect based on the same basic
principle of blue-shifted light pressure that provides
both the pumping and positive feedback, 
can be observed using all kind of particles,
from single atom to a macro-object, e. g. submillimeter
dielectric sphere, as long as the right conditions,
in particular good optical resonance, are satisfied.

The "one-atom" maser [2] and laser [3] generate
optical photons with quantum-mechanical statistics [4],
whereas a single particle $motional$ SSO
allows for highly excited \emph{classical} motion
and interesting applications.
While trapped/cooled ions, atoms [5-20] and molecules [21]
have became fascinating objects of research
on their quantum properties and applications [5-10],
their "classical" dynamics remains somewhat less explored.
In this Letter, we propose to use a trapped
atom, ion, molecule, or macro-particle
to excite a \emph{motional} SSO powered by
Doppler-affected radiation
pressure of light blue-detuned from an atomic absorption line.
This Light Activated Self-Sustained Oscillator (LASSO) is
a controllable and stable system with
large amplitude excitation and all the major SSO features;
for certain conditions it is almost thresholdless.
A LASSO could be used 
as a rotation sensor based on the "Foucault pendulum" in a trap,
single atom spectroscopy, mass-spectroscopy, isotope separation, etc.
The idea of using near-resonant $red-detuned$ light to impose
strong damping on the motion
of atoms and ions via Doppler-affected radiation pressure [16-20]
provided a powerful tool for laser cooling [5-15].
On the other hand, it became obvious long ago [22] that
a \emph{blue-detuned} laser would facilitate
a Doppler instability and the positive feedback needed for SSO.
For a particle in a trap, the required
laser intensities are extremely low and allow
for \emph{cw} operation of a LASSO,
providing for interesting applications.
Blue-detuned radiation was proposed also for use 
in atom waveguides and concave traps [23].

\section{Forces and equation of motion}
Let us consider the simplest model of a
classical particle motion in a \emph{1-D} harmonic
potential in the $z$-axis
with a frequency $\Omega$ under the action
of EM-waves counterpropagating
in the same $z$-axis, whereby
it is governed by the equation:
\begin{equation}
\ddot{z} +  \Omega^2 z = 
[ F_L ( z , \dot{z} , t ) + 
F_T ( z , \dot{z} ) ] {/} M ;
\label{1}
\end{equation}
where the "dot" designates $d / dt$,
$z$ is the atom location,
$F_T$ is a damping force due to losses in a trap,
$F_L$ is a light-pressure force, and $M$ is a particle mass.
A common damping factor is a drag force
due to residual rarefied gas (see e. g. [24]).
In the approximation of the so called
\emph{free  molecule flow}
it can be roughly estimated by simply computing
the collisions of a moving micro- or macro-particle
with a low-temperature molecules of the gas:
$F_T^{(g)} / M \approx  -
\dot{z} | \dot{z} | / L_g$,
where $L_g = $ $( M / M_g + 1 ) \cdot ( N_g \sigma )^{-1} / 2$
is the mean free path, $ ( N_g \sigma )^{-1}$,
scaled by a mass-factor due to
energy/momentum transfer
to a gas molecule of mass $M_g$;
$N_g$ is the gas number density,
and $\sigma = \pi ( d_p + d_g  )^2 / 4$
is the cross-section of
a collision between a particle and a gas molecule
of respective diameters $d_p$ and $d_g$,
assuming that they both are ideal spheres.
In a static (ion) trap, on the other hand, 
an another factor is the energy decay of a charged particle
via lossy trap circuits [10];
in this case $F_T^ {(t)} / M = - 2 \Gamma \dot{z}$,
where $( 2 \Gamma )^{-1}$
is a relevant relaxation time.

The radiation force $F_L$ is caused by laser beams with
their frequency $\omega$ in the laboratory frame
being blue-detuned from an atomic frequency $\omega_A$.
Since in most of configurations of interest,
the pumping beams can be assumed weakly focused
and regarded as plane waves,
the scattering force due to photon absorption [5-9,16-20]
(to be followed by spontaneous emission) is dominant
over the gradient (or dipole, or stimulated emission)
force [5-9,11-15,23], which is due to spatial
inhomogeneity of each wave [25].
We consider here only the spontaneous component of $F_L$;
while simplifying the basic theory of LASSO,
this assumption is not critical.
Assigning the subscript "+" to a wave
propagating in the positive $z$ direction,
and "-" to the opposite direction, one has
$F_L = F_{+} - F_{-}$, where
$F_{\pm} = \hbar k  ( d N_{\pm} / d t ) $,
$k  = \omega  / c$,
and $d N_{\pm} / d t$ is the rate of photon absorption
by the atom from the respective waves.
In the case of large detuning
(see below), for most part of a motional cycle,
the absorption/radiation is a virtual
transition to be regarded an instantaneous elastic
scattering, leaving the excited level depopulated.
We also assume through the paper
that in the case of single atoms and ions,
the trap temperature is kept significantly low,
which is a common situation in most of the trap experiments
with micro-particles.
In application to a two-level model in the case of micro-particles,
we also make a realistic assumption that the trapping potential 
is the same for both the ground state and the excited state, 
i.~e. that the resonance frequency is the same at any point of the trap. 

The atom energy losses in vacuum
(and hence the required laser pumping to overcome it) are low,
and all the major effects emerge at intensities many orders of
magnitude lower than the saturation intensity;
and typically (and preferably) the trap frequency $\Omega$ is much
lower than the atomic absorption linewidth, $\gamma$
(the "weak binding" limit).
Thus, one can use a no-saturation and
"polarization-follows-the-driving" approximation,
especially for large detunings
$\Delta \omega = \omega  - \omega_A$,
when $\Omega^2 \ll \gamma^2 +  \Delta \omega^2$,
which is the case of most interest;
its results coincide with those of a
classical Lorentz absorption model.
The instantaneous (on the motional scale)
radiation forces are then:
\begin{equation}
F_{\pm} =  ( \hbar k  ) \cdot  ( \gamma \Omega_R^2 )
{[ \gamma^2 + {( \Delta \omega 
\mp k  \dot{z} )}^2 ]}^{-1}
\label{2}
\end{equation}
where $\mp k  \dot{z} ( t )$ are instantaneous Doppler
shifts of atomic frequency with regard to
the "$\pm$" waves respectively
(note that for large oscillations, the peak shifts,
occurring near the center of oscillations, $z = 0$,
are large, $k  | \dot{z}_{pk} | \gg \gamma$),
$\Omega_R  = e \vect{d} \cdot \vect{E}  / \hbar$ and
$\vect{E}$ are the
Rabi frequency and the amplitude of each wave,
and $e \vect{d}$ is an atomic dipole moment.
For small oscillations,
the force $F_L$ can be written as
\begin{equation}
F_L \equiv F_+ - F_- \approx  \dot{z} \cdot \Delta \omega \cdot Q , \ \ \ \
where \ \ \ \
Q = \hbar k ^2 \cdot \gamma
\Omega_R^2 / ( \gamma^2 +  \Delta \omega^2 )^2 > 0
\label{3}
\end{equation}
If $\Delta \omega > 0$,  one has $F_L / \dot{z} > 0$,
which makes $F_L$ an anti-damping force.
It is easily understood from the photon absorption viewpoint:
as opposite to the atom cooling by red-shifted photons,
when the blue-shifted photon is absorbed by an atom
at its resonant frequency, the excess energy
of the absorbed photon would go into kinetic energy
of the atom and heat it up.
If $F_L / M$ overcomes the damping $2 \dot{z} \Gamma$ in Eq. (1),
the oscillations build up resulting in SSO,
which is essentially a classical "squeezed"
process with well determined, low-fluctuation amplitude
and uncertain phase.
The nonlinearity of $F_L$ \emph{vs}
$\dot{z}$ at some point arrests this growth,
and a limit cycle (a steady-state mode of SSO) is established.

The motion of a particle attaining up to $\sim 1 eV$
energy typical for a static trap for ions,
corresponds to $\sim 10^4 K$ temperature.
However, although the process in consideration 
can be viewed as something
opposite to cooling, it should not be
confused with simply heating the particle:
if the pumping is well above threshold,
its energy is transformed into
a highly ordered, coherent SSO motion, which 
differs from a thermal heating  the same way
as the sound of violin differs from a street noise,
or laser radiation -- from that of a black-body radiation.

Using envelope approximation [1]
(applicable since $\Gamma + a \Omega / \tilde{L} \ll \Omega$),
i. e. $ z \approx a \sin ( \Omega t  + \phi )$,
where $a ( t )$ and $\phi ( t )$ are
slowly varying amplitude and phase
of the oscillations respectively,
one arrives at the equation for the
dynamics of the peak velocity,
$v ( t ) = \dot{z}_{pk} = a ( t ) \Omega$, alone:
\begin{equation}
\dot{v}  = v \cdot [ G  ( v^2 ) - \Gamma -
| v | / \tilde{L} ] ,  \ \ \ \ \
G = ( 2 \pi M v^2 )^{-1}  \int_{- \pi}^{\pi}
{F_L ( \dot{z} ) \dot{z} d ( \Omega t )}
\label{4}
\end{equation}
where $G$ is the gain due to the radiation force
averaged over the oscillation cycle,
and $\tilde{L} = L_g ( 3 \pi / 4 )$;
all the term in $rhs$ of Eq. (4)
are the Fourier $\Omega$-components of
$ F_L ( t ) / M$, $ F_T ^ {(t)} ( t ) / M$,
and $ F_T ^ {(g)} ( t ) / M$, respectively.
Using dimensionless parameters $\delta$, 
$\rho $, and $u$, defined as
\begin{equation}
\delta = \Delta \omega / \gamma ; \ \
\rho = \Omega_R / \gamma =
| e \vect{d} \cdot \vect{E} |  / \hbar \gamma ;~~
u = v k / \gamma ,
\label{5}
\end{equation}
one evaluates an integral for the nonlinear gain $G$ in Eq. (4) 
resulting in:
\begin{equation}
G ( \rho , \delta , u^2 ) = 
{ 2 ^{3/2} ( \hbar \omega^2 / M c^2 ) \delta \rho^2 }
/ D ( \delta , u^2 )
\label{6}
\end{equation}
where
$D = C ( A^2 - B u^2 + A C )^{1/2} $ 
is a nonlinear dispersion factor, with
$ A = \rho_{sat}^2 = 1 + \delta^2$
-- a normalized saturation  intensity,
$B = \delta^2 - 1$, and $C = ( A^2 - 2 B u^2 + u^4 )^{1/2}$.

\section{Stationary self-sustained oscillations}
The \emph{cw} mode follows from
(4) with $\dot{v}  = \dot{u} = 0$.
One of \emph{cw} solutions is $u = 0$.
The \emph{cw} mode with $u \neq 0$ is due to
the losses being exactly compensated
for by the light induced gain,
$G_{u \neq 0} = \Gamma + | u | \gamma / k L $,
and the motional amplitude $u$
is determined implicitly by the equation:
\begin{equation}
\rho^2 = 2 ^{ -3/2} ( \rho_T ^ 2 + \rho_g^2 | u | )
D ( \delta , u^2 ) / \delta
\label{7}
\end{equation}
with "trapping" and "collisional" loss parameters:
\begin{equation}
\rho_T ^2 = 
( M c^2 / \hbar  \omega ) ( \Gamma /  \omega ) , \ \
\rho_g^2 = \rho_T^2 / r ; 
\label{8}
\end{equation}
and their ratio $r = ( \Gamma / \gamma ) k \tilde{L}$.
The parameters $\rho_T^2$ and $\rho_g^2$ are 
tremendously lower than the saturation intensity,
$\rho_{sat}^2$.
Using as an example the \emph{static} (ion) trapping
a $Na$-like $^{24} Mg^{+}$ ion [10]
with $\lambda \sim 280 nm$,
$d \sim 1.5 \times a_0$,
where $a_0$ is the Bohr radius,
$\gamma \sim 1.2 \times 10^8 s^{-1}$,
gas of $H_2$ with
$d_p + d_g \sim 0.2 nm$,
$\Gamma \sim 10^{-4} s^{-1} $ and a
pressure of $7.7 \times 10^{-9} torr$, one has
$\rho_T^2 \sim 4 \cdot 10^{-14}$,
and $\rho_g^2$ $\sim 1.2 \times 10^{-15}$.
The threshold pumping, $\rho_{thr}^2$,
required to excite the
LASSO in a "soft" way, i. e. from zero,
$G_{u = 0 } =  \Gamma$, is due only to $\rho_T$:
\begin{equation}
\rho_{thr}^2 =  \rho_T^2 
{ ( 1 + \delta^2 )^2} / 2 {\delta} .
\label{9}
\end{equation}
It is the lowest, $\rho_{min}^2 =  
( 2 /  \sqrt{3})^3 \rho_T^2$,
at $\delta = $ $3^{-1/2}$, and corresponds to
the field intensity 
$\rho_{min}^2 ( \hbar \gamma / e d )^2 ( 240 \pi )^{-1}
\sim 10^{-14}$ $W / cm^2$,
so that the LASSO is virtually $thresholdless$
(but may not be so for large detunings).
The rest of the LASSO characteristics depend on the ratio
$r \equiv \rho_T^2 / \rho_g^2$.
In our example $r \sim 34 \gg 1$,
as is typical for ion trapping.
One can then neglect collisional losses;
based on Eq. (7), Fig. 1 depicts the motional
amplitude $u$ in such a case
$vs$ the normalized laser intensity, 
$I = \rho^2 / \rho_{min}^2 $.
\begin{figure}
\begin{center}
\includegraphics{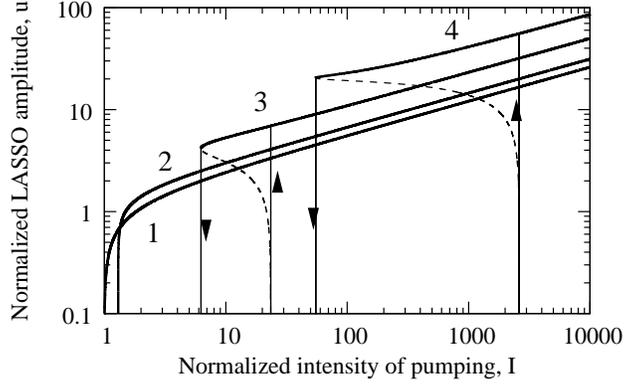}
\caption{The normalized LASSO amplitude (and peak velocity) $u$
\emph{vs} the normalized pumping, $I = \rho^2 / \rho_T^2$,
for $r \gg 1$.  Curves: (a), 1: $\delta = 1 / \sqrt{3}$,
2: $\delta = 1$, 3: $\delta = 4$, 4: $\delta = 20$.
Solid lines - stable, broken - unstable modes.
Arrows show the directions of hysteresis jumps.}
\end{center}
\end{figure}

\section{Multi-stable \textit{cw} modes and hysteresis}
In general, parameter $r$ may vary widely
(e. g.  for neutral particle trapping, $r \ll 1$,
under proper conditions, see below),
which affects the dependence of
the LASSO amplitude on the pumping intensity.
However, the common and most remarkable
feature emerging here for any $r$, is large hysteresises
when the detuning and pumping are sufficiently large.
The hysteresis-free domain is larger
for the case of dominant collisional losses,
$r \ll 1$, since those losses
are negligibly low at small amplitudes;
in this case the hysteresis-free
LASSO excitation exists for $0 < \delta < 2.75$,
for any $\rho$, and for $\rho^2 > 16 \rho_g^2$,
for $\delta > 2.75$.
If $r \gg 1$, i. e. dominant trapping losses, 
a similar domain is $0 < \delta  < 1$ for any $\rho$.
At the onset of hysteresis
$u \approx 2.15$ if $r \ll 1$,
whereas $u \ll 1$ if $r \gg 1$ (as in the above example).
Within a LASSO hysteresis loop,
there are two to three non-zero
steady-states, depending on $r$.
Examination of particle dynamics using Eq. (4) shows that
the limit cycle with the largest $u$ is always stable.
The same is true for the lowest  $u$ in the case
of three non-zero $u$'s (emerging at $r \ll 1$),
whereas the intermediate limit cycle is unstable.
With two non-zero $u$'s (emerging at $r \gg 1$),
the limit cycle with lower $u$ is unstable.
The zero-point steady-state, $u = 0$,
is always stable at $\rho^2 < \rho_{thr}^2$,
and unstable otherwise.
The solution for the lowest branch,
$u_{low}^2 \ll 1 + \delta^2$,
away from the immediate vicinity of a hysteresis jump, is
\begin{equation}
u_{low} \approx r 
( \rho^2 / \rho_{thr}^2 - 1 ) .
\label{10}
\end{equation}
On the upper branch,
$u_{high}^2 \gg 1 + \delta^2$,
in the limit $\rho_T^2 \ll u \rho_g^2 $,
one has
$u_{high} \approx ( 2 \delta  \rho^2 / \rho_g^2 )^{1/4}$,
while in the limit
$\rho_T^2 \gg u \rho_g^2$,
$u_{high} \approx ( 2 \delta \rho^2 /$ 
$ \rho_T^2 )^{1/3} =
( 2 \delta I )^{1/3} \sqrt{3} / 2 $.
The former limit is best seen on the
\emph{log-log} plots in Fig. \ 1.
The higher the magnitude of $\delta$ and $\rho^2$,
the larger is the loop.
Evaluating the "contrast of hysteresis"
as the ratio $w_{hys}$
of the "up-jump" intensity, $\rho_{up}^2$,
to the "down-jump" intensity, $\rho_{dwn}^2$,
and making use of Eq. (7) for $\delta^2 \gg 1$,
we estimate that at the down-jump
$u \sim \delta$ regardless of $r$.
Thus, if $r \gg 1$,
$\rho_{dwn}^2 \sim \rho_T^2 \delta^{3/2}$,
while if $r \ll 1$, $ \rho_{dwn}^2 \sim \rho_g^2 \delta^{5/2}$.
The intensity at which the up-jump
occurs is $\rho_{up}^2 \approx
\rho_{thr}^2$ if $r \gg 1$.
In the limit $r \ll 1$, one obtains
$u \sim \delta / 2$ at the onset of the up-jump.
Hence $2 \rho_{up}^2 \sim 3^{3/2} \rho_g^2 ( \delta / 2 )^4$.
Thus, in both limits,
\begin{equation}
w_{hys} \approx const \cdot \delta^{3/2} , \ \
const = O ( 1 ) .
\label{11}
\end{equation}
Here $const \sim 0.5$ if $r \gg 1$, and $\sim  0.16$,
if $r \ll 1$.
Typically, if $\delta \gg 1$,
the contrast $w_{hys}$ is large.

To explain the LASSO hysteresis, we note
that the larger the detuning,
the lower is the dispersion, hence
the smaller Doppler \emph{positive feedback}
for \emph{small} oscillations.
However, sufficiently \emph{large} oscillations
at the \emph{same} pumping can be self-sustained,
provided their peak Doppler shift $u$
brings the atom closer to an \emph{exact} resonance
where the dispersion is the strongest,
i. e. $| u | \sim \delta$ or
$ k  | v | \sim \Delta \omega $, see Eq. (6).
Thus, once large oscillations are excited,
they can be supported even by a \emph{lower} pumping intensity.
If $u , \delta \gg 1$,
a peak positive feedback occurs at the instance
when the Doppler effect Eq. (2) in \emph{one of the  waves}
compensates the detuning,
$ k  \dot{z} ( t ) \sim \pm \Delta \omega$.
This is reminiscent of the game of tennis:
it can be sustained
only if the ball speed is high enough.
(Note here that the LASSO hysteresis
differs from that in cyclotron excitation
of a single electron [30-32], since in the latter case one
is dealing with $driven$ and not SSO motion.)
      
Both the pumping detuning $\delta$ and intensity
$\rho^2$ are easily controlled.
Even at the extremes,
the LASSO operates far below the saturation of the
atomic absorption.
The weak binding, $\Omega / \gamma \ll 1$
is most favorable for LASSO operation.
(A strong binding will be considered by us elsewhere.)
Typically, the SSO amplitude is much larger
than the laser wavelength;
thus the standing wave pattern [24]
in $F_L$ was justifiably neglected.
It is also worth noting that while
standing wave wave was used here
for simplicity sake, actually the same
LASSO effect can attained by using only 
one traveling wave [25]; 
the main difference is that in \emph{cw} mode 
in this case the center of particle
motion would not coincide with the
lowest point of trap potential.

For large LASSO oscillations,
the harmonicity of the trap potential in Eq. (1) would not hold,
thus affecting the LASSO frequency.
If the anharmonicity is symmetric, the term
$\Omega^2 z$ in Eq. (1) could be replaced
by $\Omega^2 z ( 1 \pm z^2 / z_T^2 )$,
where $z_T$ is the half-size of the trap;
signs "-" and "+" correspond
to "soft" and "hard" potentials, respectively,
and the LASSO frequency is $\Omega_{NL} \approx \Omega
({1 \pm 3 a^2 / 8 z_T^2 })$,
provided $a^2 \ll z_T^2$.
A motional quantum excitation number,
$( u^2 / 2 ) (  \gamma / \omega )^2 M c^2 / \hbar \Omega$,
for typical conditions is $10^5 - 10^7$;
thus, the system may be regarded as strongly classical
at low frequencies.
However, at the "optical end",
under certain conditions
it may become strongly quantum, since the 
absorption of photons by the atom,
and resulting spontaneous emission of photons
in all possible direction may result 
in the spontaneous deflection of atom 
from a classically-prescribed trajectory,
which may be regarded as a broad-band noise
induced by optical pumping.
Near the threshold of oscillations,
this noise will result in the broadening
of the spectral line of oscillations,
similar to any other self-sustained
oscillator, such as e.~g. shot noise
in electronic SSO, or spontaneous
radiation in laser near the
threshold of its excitation.
However, similarly to those system,
one should expect that as the pumping 
significantly exceeds that threshold,
$\rho^2 \gg \rho_{thr}^2$,
the spontaneous noise becomes insignificant, and
the oscillations will be strongly coherent
and their linewidth -- drastically reduced. 

An elliptic or circular LASSO orbit
in a \emph{2-D} LASSO can be attained by using
two pairs of counterpropagating laser beams,
with their axes normal to each other;
circular motion will require the same pairs' intensities.
One of the differences of such a "LASSOtron" resonance
from a magnetic cyclotron resonance 
and its hysteresises [26-28]
is that the particle does not have only one direction of revolution
prescribed by a \emph{dc} magnetic field and its charge,
but may instead, depending on initial conditions,
revolve now in any direction in the plane of the beams.

\section{Traps and settings}
The experimental conditions for LASSO observation
can be arranged based on the existing traps.
In fact, the heating observed in [29,30]
may indicate a $transient$ SSO excited
during that part of the cycle of the $driven$
$rf$ side-band motional oscillations in the Paul
trap, when due to \emph{driven} Doppler shift,
an ion sees the laser (on average red-detuned) as blue-detuned.
However, the related micromotion is
far from a well defined steady-state LASSO regime.
The strong binding arrangement in [29,30]
should produce a much more complicated picture of motion than LASSO.
The hysteresis in the motion of a
trapped single ion has been observed
experimentally in [31] and numerically in [32].
The major nonlinearity in [31,32] comes however from
$two-photon$ excitation of a \emph{three-level} ion
(e. g. $Ba^+$) pumped by \emph{two}
lasers with their frequencies near-resonant
to different atomic transitions.
The LASSO, on the other hand,
involves the simplest, two-level atom model
with a single-photon resonance, and is based only on
the Doppler effect and not the \emph{atomic} nonlinearity.
The LASSO produces huge oscillations that take
the system far beyond the Lamb-Dicke limit,
$k a \ll 1$, common in trapping/cooling
physics, as well as huge hysteresises
with a contrast of few orders of magnitude.

The issue of the most appropriate static (ion) trap
for the LASSO experiment has to be considered 
separately, with recent developments in the field
offering a wide range of possible traps [37].
The simplest way to observe LASSO could be to use the
Penning trap, with the particle self-sustained oscillation
exited  along the axis of symmetry of the trap,
i. e. between the trap caps and normally to the ring electrode.
The main condition then is that the cyclotron
frequency of a particle in a strong \emph{dc} magnetic field
be much greater than the vibrational LASSO frequency
$\Omega$, but this is typical for Penning traps [10,37].
This arrangement can be used for mass-spectroscopy
and isotope separation; however it cannot
support Foucault oscillator, since the LASSO motion 
here has strongly preferred oscillation axis.
On the other hand, the Paul trap
makes it feasible for a particle to 
choose its main direction of oscillations
in a plane normal to the trap axis,
and thus it could be appropriate for realization
of \emph{2-D} Foucault LASSO oscillator, although a \emph{1-D}
model implied by Eq. (1)
will be a rough approximation to the full motion.

While a single-atom experiment is of fundamental
significance, for a proof-of-principle experiment
a charged macro-particle in the Penning
of Paul trap is perhaps a better, and much more
attainable and controlled candidate.
(Notice, e.~g. the use of an aluminum dust
to demonstrate trapping in many experiments [37].)
A cluster of e. g.  alkali or rare-gas atoms,
a dielectric or metallic charged sphere,
or even a charged oil droplet, similar to the
ones used by Millikan in his experiments
on the measurement of an electron charge,
may become an easy setup with a simple static trap
for a charged particle and a low-power laser.
In this case, for the pumping exceeding
the threshold by a few times,
the entire situation is purely classical.

While static traps allow
for ion trapping [5-9,10,37] only, a neutral atom can
be trapped optically by gradient (dipole) forces [11].
It is attracted to the high-intensity areas
if $\Delta \omega < 0$, and the low-intensity
areas if $\Delta \omega > 0$.
An all-optical LASSO is attained
by gradient-trapping an atom $Na$, with
$\lambda \sim 590 nm$ on the $3s \rightarrow 3p$ D-line,
$d \sim 2.2 \cdot a_0$, and
$\gamma \sim 2.7 \times 10^7 s^{-1}$,
in the focal area of a single red-detuned laser [12],
while LASSO-pumping (using scattering forces)
is attained by an a weakly focused blue-detuned laser.
The trapping laser here induces
also a Doppler spontaneous damping.
To use a blue-detuned laser only,
one can arrange two collinear counterpropagating beams
having their foci set apart [13] and
a "doughnut" [33,34] profile with non-zero intensity on the axis.
The only damping here is due to collisions,
and LASSO here is completely thresholdless.

In a future research, it may also be worth considering the case 
of an atom submitted to both a blue and a red-detuned laser, 
so as to have at the same time damping by Doppler cooling 
and a LASSO effect.  
By using different detunings, powers, and transitions 
(different linewidth), it might be possible to find 
a regime where the velocity-dependence of the damping 
and the LASSO forces are different enough 
so as to generate stable oscillations.

\section{"Foucault oscillator" and other feasible applications}
The LASSO has an interesting potential as a rotation sensor
based on a "Foucault pendulum" (FP) effect: 
a LASSO-excited (and then left alone) particle
conserves its oscillation direction in space.
Along with a gyroscope and an EM Sagnac effects,
the Foucault pendulum [35] is one of very few effects/devices,
which unequivocally demonstrate that the universe affects
our "regular" mechanics and electrodynamics.
However, unlike the other two effects,
the Foucault pendulum needs a cathedral-size
structure for its realization, which severely
limits its applications for e. g. rotation detection
and inertial navigation; 
besides, it relies on the gravitation,
which imposes even stronger limits on these applications.
The need in very long pendulum string
comes out of the requirement for
a sufficiently long relaxation time.
Lowering the frequency of the 
pendulum makes this time longer even for the same finesse
of the system; furthermore, the damping per cycle
is lower for higher ratio of the string length
to its diameter; hence the long pendulum.
The gravitation/acceleration environment simply makes 
the pendulum possible (for example,
the Foucault pendulum -- or any pendulum -- 
would be impossible in the space).
Both these drawbacks of the Foucault pendulum can be overcame
by using a single particle trapped in a symmetrical 
\emph{3-D} or \emph{2-D} trap in a vacuum.
This would allow one to have a small-size
motional oscillator with very low damping 
(which can further be completely eliminated 
by a low-intrusive positive feedback provided 
by a few laser beams, see below), whose oscillation and
momentum-conservation properties will be "portable"
in the sense that they be provided by the
trap potential, and not by the local gravity.
All this may put the LASSO used as a "Foucault trap oscillator" 
into the realm of inertial navigation
applications similar to those of mechanical and laser 
(i. e. Sagnac-based) gyroscopes.

For the FP observation in the \emph{2-D} configuration,
one can use two degenerate degrees
of freedom found in some of existing trap,
to make a "Foucault plane".
Small imperfections, $\Delta U$, in the 
\emph{2-D} symmetry of that
potential can be made negligible by proper design and machining,
as well as by choosing a heavy particle and a large LASSO amplitude,
similarly to a regular FP.
To this end, macro-particles offer the best possibility.
Similarly to [11], 
to optically exert a light-pressure on a macro-particle,
one may use very narrow Mie-Debay resonances [36]
of dielectric sphere, which are
due to the coupling of the laser to the high-finesse
whispering gallery modes.
The critical (lowest) rotational rate picked up by the FP, is
$\Omega_{cr} \sim a^{-1} ( 2 e \Delta U / M )^{1/2}$.
If $M = 1  g$, $a = 1  cm$, and $\Delta U \sim 10^{-5} V$,
one has $\Omega_{cr} / 2 \pi \sim 10^{-9} Hz \sim
10^{-4}$ of the earth rotation rate, which is
sufficient for inertial navigation.

A slowly dissipating particle energy
can repeatedly be replenished by a laser aligned to the
new direction of oscillations in the lab frame.
As was noted above, in principle, the random character 
of the exciting force (spontaneous photons emitted in all 
directions) may cause momentum diffusion, including in the 
plane perpendicular to the oscillation.
However this might be a problem only in the case 
of micro-particle, and only near the threshold 
of the LASSO-oscillations; with increased
pumping, the ratio fluctuation/amplitude will go down.
And this factor can be completely ignored
in the case of macro-particle, as in the above example.
No regular classical Foucault pendulum even in non-rarefied air 
has ever been known for quantum-induced momentum diffusion.

The LASSO can be used for single particle spectroscopy 
by studying LASSO amplitude \emph{vs} laser frequency.
Another related application is isotope separation,
whereby a laser tuned in between atomic lines of two isotopes,
cools down the isotope with higher atomic frequency
and LASSO-excites the other isotope,
pushing it out if the pumping is sufficient.
This approach can be modified and enhanced by using hysteresis and
the fact that the up-jump intensity of
the hysteresis is very sensitive to the detuning.
With a laser blue-detuned from \emph{both} the atomic frequencies,
one can attain large LASSO-excitation of one of the species while
keeping the other one at rest.
Gradient force isotope separation by a focused Gaussian laser beam
[14] could also be greatly enhanced by Doppler damping (instability),
which is a significant factor in pulling in (pushing out)
higher (lower) atomic frequency species.

\section{Conclusion}
In conclusion, weak \emph{cw} laser radiation can induce
a Doppler instability and large self-sustained motional oscillations
of a trapped single particle via
the radiation pressure of light blue-detuned
from an absorption resonance.
The oscillations exhibit huge hysteresis
and may easily reach a trap-size orbit.
The effect can potentially be used for
rotation sensing, inertial navigation, single-atom spectroscopy,
mass-spectroscopy, and isotope separation.

This work is supported by AFOSR.
\end{document}